\documentclass[aps,prl,english,showpacs,twocolumn]{revtex4}
\usepackage{latexsym}
\usepackage{bm}
\usepackage{babel}
\usepackage[T1]{fontenc}
\usepackage[latin1]{inputenc}
\usepackage{mathrsfs}
\usepackage{amsmath, amssymb}
\usepackage{graphicx}
\usepackage{ae}
\usepackage{natbib}

\begin{document}
\title{Vibrational excitation of diatomic
molecular ions in strong-field ionization of diatomic molecules}
\author{Thomas K. Kjeldsen}
\author{Lars Bojer Madsen}
\affiliation{Department of Physics and Astronomy,
  University of Aarhus, 8000 \AA rhus C, Denmark}

\begin{abstract}
A model based on the strong-field and Born-Oppenheimer approximations
qualitatively describes the distribution over vibrational states
formed in a diatomic molecular ion following ionization of the neutral
molecule by intense laser pulses. Good agreement is found with a
recent experiment [X. Urbain {\it et al.}, Phys. Rev. Lett. {\bf 92},
163004 (2004)]. In particular, the observed deviation from a
Franck-Condon-like distribution is reproduced. Additionally, we
demonstrate control of the vibrational distribution by a variation of
the peak intensity or a change of frequency of the laser pulse.

\end{abstract}
%33.80.Eh Autoionization, photoionization, and photodetachment
%33.80.Rv Multiphoton ionization and excitation to highly
%excited states (e.g., Rydberg states)

\maketitle

Intense investigations during the last decades have led to quite a
detailed understanding of the interaction between strong laser
fields and atoms (see~\cite{Becker05} for a review). This includes,
e.g., the process of high-harmonic generation which is by now a
valuable source for the production of coherent ultraviolet light.
For molecules, however, the extra degrees of freedom introduced by
the presence of more than one nucleus lead to a much more involved
picture, and the description of the strong-field driving of such
systems far from equilibrium is still a challenge in theoretical
physics (for reviews see~\cite{Bandrauk,Posthumus}).

The main topic of this work is to investigate the partitioning of
energy among the electronic and nuclear degrees of freedom when
ionizing a diatomic molecule by a strong laser field. This is a topic
of much current interest: until recently it was assumed that the
distribution over vibrational states formed in the molecular ion
following ionization of the neutral molecule follows a Franck-Condon
(FC) distribution, and this assumption is still frequently used
(see~\cite{urbain04} and references therein). A recent experiment,
however, reported a non-FC distribution and suggested that the use of
the FC principle is inaccurate because the rate for tunneling
ionization increases sharply with internuclear
distance~\cite{urbain04}. We will show that the rapid response of the
electron to the laser field in fact ensures the applicability of the
FC principle for the nuclear motion at each instant of time during the
pulse. It is instead the variation of the ionization and excitation
rate with intensity in the focal volume that leads to a departure from
a conventional FC-like distribution. The theory is very versatile and
readily adapted to a wide range of diatomic molecules.

We note that as the strong field response of the molecular ion is very
dependent on the distribution over vibrational levels, it is desirable
to be able to control the latter. We show by explicit examples in
H$_2$, O$_2$ and N$_2$ that a simple change of the peak intensity of a
45 fs Gaussian pulse leads to a significant degree of control over the
final vibrational distribution, and hence holds the promise for the
production of a target of interest for state-specific studies.

The energy absorbed by a molecule in the field is distributed among
the electrons and nuclei. Ionization of molecules is accordingly
accompanied by a vibrational excitation of the nuclear motion in the
ion formed. To illustrate the accuracy of our approach, we first
consider H$_2$ for which strong-field experiments are
available~\cite{fabre03,fabre04,urbain04}. For H$_2$ both
photoelectron spectroscopy with UV light sources~\cite{turner} and
ionization by $100\, \mbox{eV}$ electron impact \cite{urbain04}
resulted in vibrational distributions of H$_2^+$ in accordance with
the FC principle, i.e., governed by the projection of the initial
vibrational ground state wave function $\nu' = 0$ onto the set of
vibrational states of the molecular ion, $\nu$. In recent
experiments \cite{fabre03,fabre04,urbain04}, the vibrational
distribution of H$_2^+$ was measured after ionization of H$_2$ by
intense laser fields. These experiments revealed very different
vibrational distributions than the experiments mentioned above. The
laser-induced ionization leads to a narrower distribution with
almost no population in the final vibrational levels above $\nu =
4$, and the actual shape of the distribution depends on the peak
laser intensity and wavelength. At low intensities, the maximum
population is observed in the $\nu = 0$ state while the FC principle
predicts a maximum population in the $\nu = 2$ state. These results
can be partly reproduced by a tunneling model with an electronic
binding energy which depends on the internuclear distance
\cite{fabre03,fabre04,urbain04}. However, the experimental
conditions in some of the experiments \cite{fabre04} do not
correspond to a pure tunneling regime, and a further shortcoming of
the model is that it cannot account for any variation in the
distribution with wavelength. It is therefore desirable to apply a
theory that is expected to be valid also in the  multiphoton regime.
An example of such a theory is the molecular strong-field
approximation with the inclusion of nuclear
motion~\cite{becker01,mishima04,kjeldsen05b}, and here we apply this
theory along the lines discussed in~\cite{kjeldsen05b}. This model has
previously reproduced various experimental observations quite
successfully~\cite{kjeldsen04a,kjeldsen05b}. For a
linearly polarized laser of frequency $\omega$ and periodicity $T =
2 \pi/\omega$, the rate of ionization differential in the direction
$\hat{\bm q}$ of the ejected electron, to a particular vibrational
state of the molecular ion after absorption of $n$ photons is  $
dW_{fi}/d\hat{\bm q} = 2\, \pi |A_{fi}|^2 q_n $ [atomic units, $e =
\hbar = m_e = a_0= 1$, are used throughout]. The magnitude of the
momentum $q_n$ is determined by energy conservation and $A_{fi}$ is
the transition amplitude,
\begin{equation}
  \label{eqn:Afi}
  A_{fi} = S_{fi}\, \frac{1}{T}\int_0^T \langle f(t) | V_F(t) | i(t)
  \rangle dt,
\end{equation}
where $S_{fi}$ is the FC factor corresponding to the nuclear
vibrational transition, $V_F(t)$ is the length gauge form for the
molecule-light interaction and $|i(t)\rangle$ and $|f(t)\rangle$ are
the initial and final electronic states, respectively, including
energy phases for the nuclear vibrational motion. The time scale of molecular
rotations is much slower than the typical pulse duration so that the
molecular orientation is considered as being fixed throughout the pulse. 
The time scale of the vibrational motion
on the other hand, is shorter than the typical pulse length and therefore we
treat this degree of freedom quantum mechanically.
The initial electronic state is the
highest occupied molecular orbital, and in the final state it is
assumed that the laser-electron interaction is much stronger than
the electron-molecule interaction, such that the final state of the
electron can be accurately described by a Volkov wave. In
Eq.~\eqref{eqn:Afi}, the electronic matrix element is evaluated at the
equilibrium distance of the nuclei in accordance with the
Born-Oppenheimer approximation. This allows us to isolate the
overlap between the nuclear wave functions expressed by the FC
factor. Under experimental conditions, the temperature will be
sufficiently low so that only the vibrational ground state of the
initial state is populated. We integrate over all directions of the
outgoing electron and sum over all accessible numbers of photon
absorptions to obtain the total rate of ionization to the vibrational
state considered. Finally, and this turns out to be essential, we
integrate the total rates over the Gaussian spatial and temporal profile
of the laser pulse and average over molecular orientations for direct
comparison with experimental data. The spatial integration is carried
out over the interaction region which is restricted to a small volume
around the beam waist in the experiment. It is not difficult to show
that relative signals are independent of focal spot size.

\begin{figure}
    \includegraphics[width=\columnwidth]{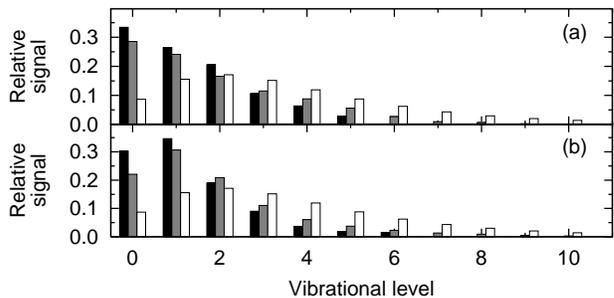}
  \caption{ The vibrational distribution of H$_2^+$ after ionization
  in an intense laser field. The black bars indicate the experimental
  observations~\cite{urbain04} and the gray bars are the predictions
  according to the present theory.  The Franck-Condon distribution is
  indicated with white bars. In both panels the laser wavelength is
  $800\, \mbox{nm}$, the pulse duration (full width at half maximum)
  is $45\, \mbox{fs}$.  The peak laser intensities are (a) $3.0\times
  10^{13}\, \mbox{W/cm}^2$ and (b) $4.8\times 10^{13}\,
  \mbox{W/cm}^2$.  } \label{fig:h2-exp} 
\end{figure}

In Fig.~\ref{fig:h2-exp}, we present the experimental and
theoretical vibrational distribution of H$_2^+$ produced by
strong-field ionization of H$_2$ for two different intensities.
For comparison also squared FC factors are shown.
At both intensities, we find good agreement between experiment and
theory.  Both distributions favor the lower vibrational states in
contrast to the FC distribution.

Our aim is to provide a widely applicable model and therefore we
simply use the field-free FC factors. These are readily available in
the literature for a wide range of molecules. In reality, the
field-free nuclear potential curves are distorted by a very intense
laser field, e.g., in H$_2^+$ this effect leads to a bond-softening
due to a coupling between the lowest $\Sigma_g$ and $\Sigma_u$
states. As the potential is changed, a new set of vibrational
eigenstates and corresponding FC factors needs to be considered. If,
however, the intensity is below $5\times 10^{13}\, \mbox{W/cm}^2$
the field-induced modification of the potential curve is minimal
\cite{urbain04} and the application of the field-free FC factors
accurate. Also, we note that dissociating channels become
increasingly important as the intensity exceeds $5\times 10^{13}\,
\mbox{W/cm}^2$, and since one might be interested in making
subsequent experiments on the {\it bound} molecular ion we will not
consider such intensities here.

Even though the FC principle is explicitly used in our theory
[see Eq.~\eqref{eqn:Afi}], the predicted and observed vibrational
distributions deviate significantly from the FC distribution. The
key concept explaining this discrepancy is the effect of channel
closings in connection with full account of the
pulse profile. By energy consideration, the number of
absorbed photons $n$ must fulfil the criterion  $n \omega =
q_n^2/2 + I_p^{\nu} + U_p$ with $q_n > 0$, $I_p^{\nu}$
the ionization potential to the level $\nu$ and $U_p =
I/(4\omega^2)$ the quiver energy of a free electron in the
laser field of intensity $I$.
The presence of $U_p$ means that the minimum number of photons needed to
reach the electronic continuum and a particular vibrational state $\nu$ in
the ion increases with intensity.
\begin{figure}
    \includegraphics[width=\columnwidth]{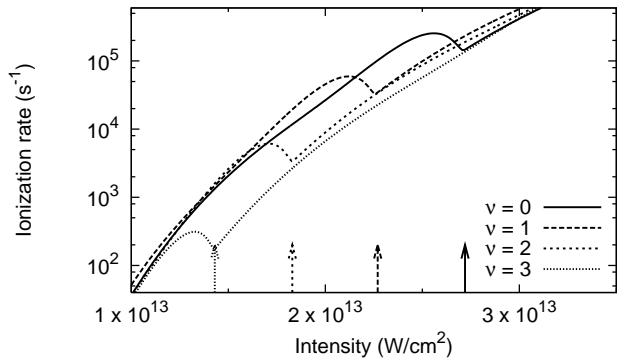}
  \caption{ Ionization rates to the lowest vibrational states of H$_2^+$
  from H$_2$ molecules aligned parallel to the laser field at the wavelength
  of $800\, \mbox{nm}$. Note logarithmic scale.
  }
  \label{fig:h2-rates}
\end{figure}
This phenomena is referred to as channel closing and the effect is illustrated in
Fig.~\ref{fig:h2-rates}, where we show the ionization rates to the
lowest vibrational levels of H$_2^+$ when ionizing molecules
aligned parallel to the laser field at a wavelength of $800\,
\mbox{nm}$. At an intensity of $1\times 10^{13}\, \mbox{W/cm}^2$ all the
vibrational states shown can be reached by absorption of 11 photons.
As the intensity increases
the thresholds shift upwards by $U_p$ and at the intensities marked by
arrows, absorption of 11 photons becomes insufficient to reach the
vibrational levels indicated.
For example, in the intensity range of $2.3-2.7 \times
10^{13} \mbox{W/cm}^2$ one needs $n \ge 12$ photons to reach the
$\nu \ge 1$ levels whereas $n = 11$ is sufficient for $\nu = 0$. Since the
rates for higher order processes ($n \ge 12$) are much lower than the rate
for $n = 11$, this explains why the $\nu = 0$ is favored by a factor of 3
over $\nu = \left\{ 1, 2 \right\}$.
Contrary, when the same
number of photons is needed to reach
all $\nu$ levels, the rate to $\nu = 0$ is generally lower than
to $\nu = \left\{ 1, 2 \right\}$ due to the smaller FC factor of the former.

The reason for finding a very different relative relationship
between the {\it signals} at the peak intensity of $3 \times
10^{13} \mbox{W/cm}^2$, Fig.~\ref{fig:h2-exp}~(a), and the {\it
rates} at the same intensity is a result of taking the pulse shape
into account. Only in the very center of the Gaussian laser beam
the intensity reaches the peak intensity. In other regions of
space the molecules are exposed to a lower intensity and
excitation to $\nu = 0$ dominates.

So far we have seen that the vibrational distribution depends markedly on the
intensity. Our next purpose is to investigate how the properties of the
laser pulse can be varied to maximize the population in a given vibrational
state. A high degree of population transfer to a definite state will be a
valuable result as it will allow for the possibility of making
state-specific experiments on the molecular ion. To our knowledge, such type
of control has not previously been explored in strong-field physics. We have
chosen to model the pulse by a simple variation of the peak intensity with
the purpose of maximizing the $\nu = 0$ population. We have performed the
optimization at the wavelengths corresponding to the fundamental- and
frequency doubled wavelengths of the Ti:Sapphire ($400\, \mbox{nm}$ and
$800\, \mbox{nm}$) and Nd:YAG ($532\, \mbox{nm}$ and $1064\, \mbox{nm}$)
lasers. During optimization the pulse duration is fixed at $45\, \mbox{fs}$
and $6\, \mbox{ns}$ for the Ti:Sapphire and Nd:YAG wavelengths,
respectively.

\begin{figure}
    \includegraphics[width=\columnwidth]{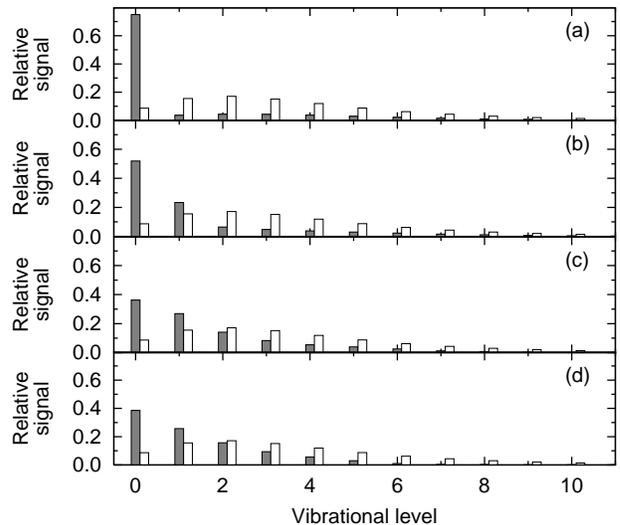}
  \caption{ The vibrational distribution of H$_2^+$ after ionization in an
  intense laser field according to the present theory (gray bars).
  The Franck-Condon distribution is indicated with white bars.
  The peak laser intensity is chosen to maximize the population in the lowest
  vibrational state.
  The laser wavelengths are (a) $400\, \mbox{nm}$, (b) $532\, \mbox{nm}$,
  (c) $800\, \mbox{nm}$, and (d) $1064\, \mbox{nm}$, and the peak
  intensities are (a) $2.0\times 10^{12}\, \mbox{W/cm}^2$, (b) $2.7\times
  10^{13}\, \mbox{W/cm}^2$, (c) $2.6\times 10^{13}\, \mbox{W/cm}^2$, and
  (d) $1.9\times 10^{13}\, \mbox{W/cm}^2$.
  The pulse duration is $45\, \mbox{fs}$ in panels~(a) and (c), and $6\,
  \mbox{ns}$ in panels~(b) and (d).
  }
  \label{fig:h2-opt}
\end{figure}
The result of the intensity optimization is shown in
Fig.~\ref{fig:h2-opt}. At the shortest wavelength, $400\,
\mbox{nm}$, a relative population as large as 75\% is produced in
the vibrational ground state at an intensity of $2\times 10^{12}\,
\mbox{W/cm}^2$. At lower intensities we can get nearly $100\%$ in the $\nu = 0$
state but at such low intensities that the ion yields are very
low. In our model, the reason for obtaining such a confined
distribution is that the ion yield is completely dominated by
5-photonabsorption which is only possible to the $\nu = 0$ state
and only at intensities below $4.9\times 10^{12} \mbox{W/cm}^2$.
We note that this explanation of the favoring of the $\nu = 0$
level is different from a previous suggestion offering an
explanation in terms of resonance enhanced
ionization~\cite{fabre04}. As the wavelength increases the
selection of the $\nu = 0$ becomes less efficient. At all
wavelengths the optimum intensity is relatively low and it will
always correspond to an intensity which is slightly below the
occurrence of the first channel closing. The decreasing $\nu = 0$
population with increasing wavelength was also observed
experimentally~\cite{fabre03,fabre04}, and the general phenomenon
that the largest population in the vibrational ground state is
obtained at relatively low intensities is in good agreement with
the experiments.

We note that the control obtained by changing intensity and/or
wavelength is not only the simplest way of gaining control, it is
probably also the only one. If, e.g., one considers the possibility
of coherent control by preparing either a coherent superposition of
initial vibrational states separated by the energy of one photon or
by utilizing a bichromatic field and optimizing the strengths and
relative phase of the two components of the field, one faces the
problem of a continuum of final states. With a proper choice of
amplitudes and phase it is possible to exclude the transition to one
particular final state which is characterized by the vibrational
state {\it and} the direction and energy of the outgoing electron.
However, one cannot gain control over the transition to all other
final states and hence an efficient selection cannot be made.

\begin{figure}
    \includegraphics[width=\columnwidth]{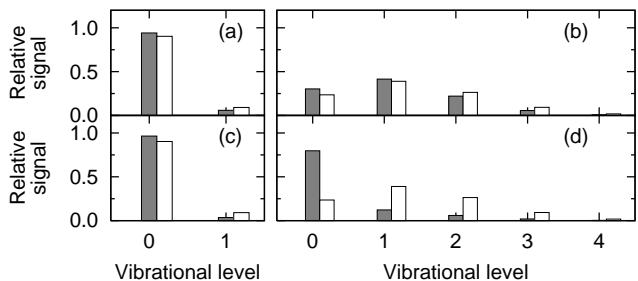}
  \caption{ The vibrational distribution of
  N$_2^+$ [(a) and (c)] and  O$_2^+$ [(b) and (d)] after ionization in an
  intense laser field according to the present theory (gray bars).
  The Franck-Condon distributions are indicated with white bars.
  In all panels the laser wavelength is $800\,
  \mbox{nm}$ and the pulse duration is $45\, \mbox{fs}$.
  A typical peak intensity of
  $5\times 10^{13}\, \mbox{W/cm}^2$ is chosen in panels (a) and (b).
  Maximization of the population in the lowest vibrational state is obtained
  with the peak intensities of (c) $2.4\times 10^{13}\, \mbox{W/cm}^2$ for
  N$_2^+$, and (d)
  $3.6\times 10^{12}\, \mbox{W/cm}^2$ for O$_2^+$. }
  \label{fig:n2-o2-opt}
\end{figure}

The vibrational wave functions of H$_2$ and H$_2^+$ lead to a
quite broad FC distribution and accordingly many vibrational
states are populated after ionization. If one considers
molecules where only a few FC factors are important, one may hope
for a more efficient optimization. To this end, and to illustrate
the wide applicability of the present formalism, we  investigated
ionization of N$_2$ and O$_2$ where the number of final
vibrational states are limited to $\nu \le 1$ and 4,
respectively, simply because the other FC factors are vanishingly
small. First, we calculated the vibrational distribution with a
laser wavelength of $800\, \mbox{nm}$, pulse duration of $45\, \mbox{fs}$
and with a typical, but
arbitrarily chosen, peak intensity of $5\times 10^{13}\,
\mbox{W/cm}^2$. We then made an optimization similar to that for
H$_2$ and the results are given in Fig.~\ref{fig:n2-o2-opt}.
The vibrational distributions obtained with a typical peak
intensity, Figs.~\ref{fig:n2-o2-opt}~(a) and (b), are quite similar
to the FC distributions. Figure~\ref{fig:n2-o2-opt}~(a) shows
typical and Fig.~\ref{fig:n2-o2-opt}~(c) optimized vibrational
distribution of N$_2^+$ after ionization of N$_2$. For N$_2$, only
two states will be populated in N$_2^+$ and we see that the FC
factors strongly favour the $\nu = 0$ state (90\%). By utilizing a
non-optimized strong laser field we obtain a $\nu = 0$ population of
94\% [Fig.~\ref{fig:n2-o2-opt}~(a)], and if we use an optimized
laser intensity we may reach a $\nu = 0$ population of 97\%
[Fig.~\ref{fig:n2-o2-opt}~(c)]. In Figs.~\ref{fig:n2-o2-opt}~(b) and
(d) we present the results for O$_2$ and we see that an efficient
selection of the $\nu = 0$ state can also be obtained for this
molecule. We find the optimal population to be 80\%
[Fig.~\ref{fig:n2-o2-opt}~(d)] compared with the FC distribution of
24\% and the non-optimized result of 30\%
[Fig.~\ref{fig:n2-o2-opt}~(b)]. When we compare the results of N$_2$
and O$_2$ with the results of H$_2$ at the same wavelength
[Fig.~\ref{fig:h2-opt}~(c)], we see that the former molecules can
indeed be brought to the $\nu = 0$ state more efficiently due to the
limited number of final vibrational states.

In conclusion, the molecular strong-field approximation with the
inclusion of nuclear vibrational motion explains the recently
observed vibrational distributions of H$_2^+$ produced by
strong-field ionization of H$_2$ molecules~\cite{urbain04}. The
theoretical and experimental distributions are both very different
from distributions predicted by Franck-Condon factors -- a result
that can be explained by the effects of channel closings in
association with the spatial and temporal extend of the laser
pulse. The theory is simple to evaluate and therefore readily
applicable to diatomic molecules. We proposed a readily available method
of optimizing the population in the $\nu = 0$ state by varying
the peak laser intensity. With this type of optimization we showed
that the vibrational distributions vary significantly with the
laser wavelength in accordance with experimental findings.
Finally, we applied the theory to  N$_2$ and O$_2$ and found that
a higher degree of control can  be obtained than in H$_2$ due to
the FC factors which are non-vanishing for fewer states in the
former cases. We note that efficient control of population
distributions will be of great value in the development of
state-specific experiments on molecular ions.

\begin{acknowledgments}
  LBM is supported by the Danish Natural Science Research Council (Grant
No. 21-03-0163).
\end{acknowledgments}

\end{document}